# Ion-Beam Radiation-Induced Eshelby Transformations: The Mean and Variance in Hydrostatic and Shear Residual Stresses

Yongchao Chen[1,2], Qing-Jie Li[1], Alex O'Brien[1], Yang Yang[3], Qi He[1], David A. Bloore[1], Joost J. Vlassak[2*] and Ju Li[1,4*]

[1] Department of Nuclear Science and Engineering, MIT, Cambridge, MA 02139, USA

[2] John A. Paulson School of Engineering and Applied Sciences, Harvard University, Cambridge, MA 02138, USA

[3] Department of Engineering Science and Mechanics & Materials Research Institute, The Pennsylvania State University, University Park, PA 16801, USA

[4] Department of Materials Science and Engineering, MIT, Cambridge, MA 02139, USA

* Corresponding authors: Joost J. Vlassak (vlassak@seas.harvard.edu) and Ju Li (liju@mit.edu)

**ABSTRACT**

Ion beam plays a pivotal role in ion implantations and the fabrication of nanostructures. However, there lacks a quantitative model to describe the residual stresses associated with the ion-beam radiation. Radiation-induced residual stress/transformation strain have been mostly recognized in the hydrostatic sub strain space. Here, we use molecular dynamics (MD) simulations to show that the response of a material to irradiation is generally anisotropic that depends on the ion-beam direction, and should be described using tensorial quantities. We demonstrate that accelerator-based ion beam irradiation, combined with the intrinsic lattice anisotropy and externally induced anisotropy (such as anisotropic mechanical loadings), causes radiation-actuated *shear* transformation strains in addition to hydrostatic expansion. We map out these complex correlations for several materials. Radiation-induced defects are shown to be responsible for residual shear stresses in the manner of Eshelby inclusion transformation. We propose such tensorial response model should be considered for accurate nanoscale fabrication using ion-beam irradiation.

1. **Introduction**

Nanomanufacturing technologies have proliferated over the past decades (1-5). Benefiting from the accurate spatial resolution down to 1 nm and flexible control of fluences and direction (6), targeted focused ion beams (FIB) have been utilized to precisely shape materials and implant dopants and defects, thereby tuning the mechanical (7, 8), electrical (9, 10), and chemical (11) properties of components. Recently, electron-beam irradiation further enables atomic engineering at the sub-1 nm scale, thus emerging as a radiation-assisted method to manipulate the atomic structure of materials (3). Beam-irradiated samples usually contain various nanoscale defect clusters (e.g., clusters of interstitials and vacancies) (2), which break the symmetry of the host lattice and show distinct shape-transformation (first-order) as well as elastic stiffness (second-order) changes in "Eshelby inclusion zones" (12) different from the perfect crystal. Such local inhomogeneities are expected to have transformation strain-volume tensors that cause residual stress field and morphology changes elsewhere.

Robust numerical predictions of radiation-induced residual stresses are needed to fully exploit the potential of ion/electron beam radiation in manufacturing. For example, optimal strain engineering (13) and nanostructure manufacturing (6, 14, 15) can be achieved if we know how to generate 3D actuation stress field through precise control of the ion beam parameters (energy and momentum of ions), to achieve "3D Printing of Eshelby Transformations (3DPET)". Such delicately controlled residual stress sources (Kanzaki force dipoles (16, 17)) can serve as the 'hundred little hands' envisioned by Feynman (18) to precisely tug on materials from near-surface region to several microns deep inside. While previous studies, by assuming isotropic fluxes or fluences of the irradiation, have shown the evolution of *hydrostatic* residual stress (14, 19, 20); the vector characteristics of a *beam* of incident ions/electrons (delta-function distribution in the ion momentum space) and possible *shear* transformation have not been systematically investigated by atomistic simulations, at the same level of scrutiny as the distribution of hydrostatic and shear residual stresses in unirradiated amorphous solids by Srolovitz et al. (21). Previous studies on irradiated materials mainly focused on the generation of compressive hydrostatic stress (19, 22-24), probably due to the complexities of the deviatoric (shear) components related to the ion-beam direction and crystal symmetry. Recently, Shao et al. (25) and Ren et al. (26) demonstrated that the scalar simplification introduces inaccuracies due to momentum monodispersity in accelerator-based ion- or electron-beam irradiation. This highlights the importance of clarifying the existence of radiation-induced deviatoric (shear) stresses.

Experimentally mapping out residual stress fields is challenging (27). Atomistic simulations are well suited to examine the cause of residual stress fields, which are the distributed Eshelby transformation strain-volume tensors. In this work, the mean and variance of hydrostatic and shear Eshelby-transformation stresses caused by ion-beam radiation are addressed. Through molecular dynamics (MD) simulations, finite element method (FEM), and statistical analysis for both cubic-symmetry metals (such as BCC iron and FCC copper) and lower-symmetry ceramic compounds such as zircon ($ZrSiO_4$), we demonstrate the existence of both average residual shear stresses and residual shear stress fluctuations under fixed-momentum irradiation (Fig. 1). These results demonstrate a complex connection between the momentum of the incident particles and the resulting average residual shear stress in various crystals, suggesting the non-scalar transformation characteristics of nanoscale irradiation cascades. We further illustrate that such average residual shear stresses originate from specific defect biases (due to momentum monodispersity/structural anisotropy) and evolve with the incident particle energy. Furthermore, we show that externally applied pre-stress (irradiation creep) is another means to tune the irradiation-induced Eshelby transformations. Such shear-transformation response to radiation could be exploited for deforming and stressing novel devices. Our work highlights the value of irradiation-induced tensorial transformation and its potential in precisely controlling nanoscale structures by 3DPET, as FIB has nm-resolution in $x$, $y$, and the depth of implantation $z$ (Bragg peak) can also be controlled by the ion energy (6).

## 2. Methods
### 2.1. MD simulations

All MD simulations are carried out using LAMMPS (Large-scale Atomic/Molecular Massively Parallel Simulator) (28). Here, we choose iron, copper, and zircon ($ZrSiO_4$) for investigation since they are common materials used in nuclear energy and in the encapsulation of highly radioactive nuclear waste. Mendelev's (29) and Mishin's (30) embedded atom method (EAM) potentials splined to the Ziegler-Biersack-Littmark (ZBL) (31) potential at small distances are adopted to describe the interatomic interaction between Fe-Fe and Cu-Cu atoms, respectively. These potentials have been tested to be appropriate for cascade events (32-34). The specific description of the potential used for zircon is consistent with former studies (35, 36). See Supplementary Note 1 for detailed validation of the interatomic potentials used in this study. The simulation box contains 10-30 iron/copper/zircon unit cells along each Cartesian dimension with the lattice constant $a$ = 2.856 Å for iron and $a$ = 3.598 Å for copper, while $a$ = 6.602 Å and $c$ = 6.093 Å for the tetragonal zircon crystal. Here periodic boundary conditions are applied to avoid the surface effects, and each simulation box is sufficiently large to fully contain the thermal spikes caused by cascade events with recoil energy of 0.2-20 keV. We dynamically adjust the time step to ensure that the maximum atomic displacement during the cascade process is limited to 0.001 Å for each time step. Prior to a cascade event, the whole system is equilibrated into a stress-free or pre-stressed state by thermalizing the system for at least 50 ps at 20 K under *NPT* conditions.

After obtaining the initial relaxed configuration, we select an existing atom as PKA or introduce a new ion, which is then assigned a specific momentum (corresponding to a specific recoil or incident energy) to initiate the cascade process. Each sample is divided into an inner and an outer region (0.3 nm in thickness). The inner system is relaxed under the *NVE* ensemble for 100 ps, while the temperature of atoms in the outer layers is kept at 20 K to dissipate excess heat from the cascade event within the inner region. To avoid statistical errors and acquire accurate tranformation distributions, for each specific setup 20-500 individual simulations are repeated by setting different random seeds for the initial atomic velocities. The specific number of trials depends on the convergent behavior, in which the standard deviation for average stresses and bond number changes should be less than 5% of the whole value. For the calculation of bond number changes, the cutoff distances for Si-O, Zr-O, etc. bonds are determined based on the first minima of their partial pair distribution functions. To study the case without momentum monodispersity, 100 non-parallel directions are obtained from the code by Raman and Yang for the Thomson problem (37). The Wigner–Seitz analysis and the Dislocation Extraction Algorithm (DXA) implemented in OVITO (38) are used to identify defects and dislocations. For the creation of an artificial vacancy cluster, we randomly remove 20% atoms in a spherical region and then equilibrate the sample. In self-ion irradiation processes, the initial positions of the incident ions are randomly adjusted within the box while maintaining sufficient distance from the existing atoms to avoid large repulsion forces.

**2.2. Finite element modeling**

The COMSOL Multiphysics software package is employed to conduct FEM analyses on stress, based on the elastic modulus of iron/zircon and residual stress data acquired from MD simulations. The crystallographic orientations of zircon are varied with artificial angles to simulate GBs. The specific setup of mechanical properties in zircon is shown in Table S2.

## 3. Results

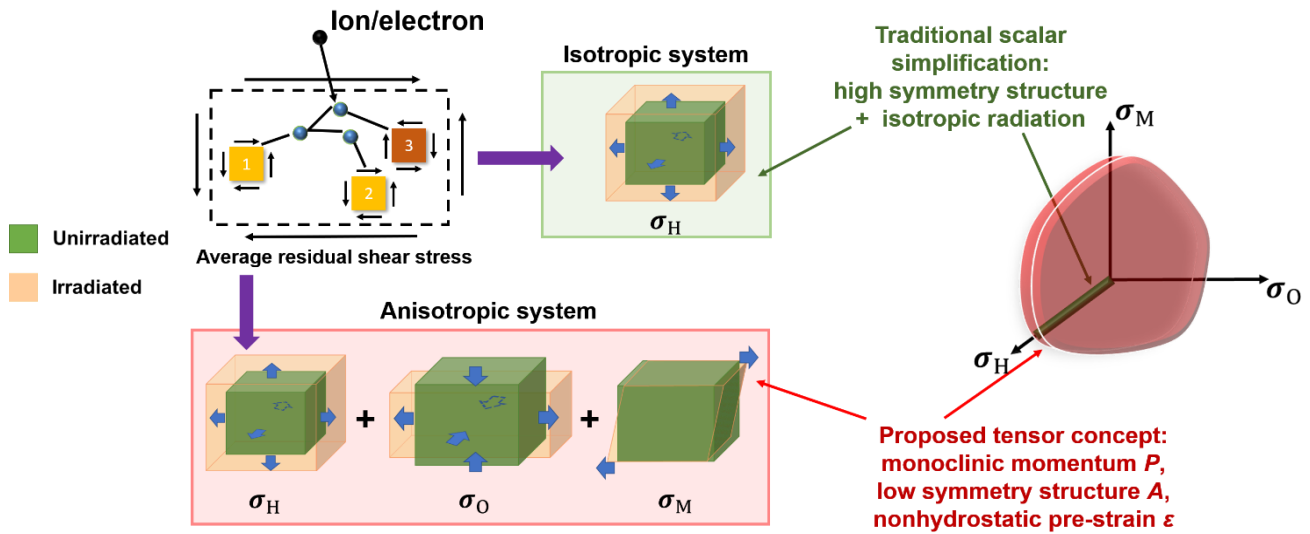

**Figure 1.** Schematic illustration of the ion/electron-beam radiation-induced residual stress and the proposed tensor concept. The whole cascade process is initiated by an incident kinetic ion/electron and propagated by created knock-on atoms (PKA, SKA, TKA, etc.). Each kinetic atom leads to one damage subcascade with corresponding residual stress, which can be divided into one hydrostatic part ($\sigma_H$) and two shear parts ($\sigma_O$, $\sigma_M$) where O stands for orthorhombic and M stands for monoclinic. The whole damage zone may penetrate a few nanometers to tens of microns, causing stress fields to overlap among different nanoscale subcascades. Individual nanoscale subcascades may together show an average hydrostatic and shear stress. For the overall post-irradiated residual stress field, an isotropic system (green colored) stimulates purely hydrostatic residual stress, a scalar quantity presented by the green rod. However, in a general space formed by several anisotropic factors (momentum, structural asymmetry, and non-hydrostatic pre-strain), radiation will cause a tensor response in the full stress state space (red-colored).

Figure 1 illustrates a general framework for radiation-induced Eshelby transformations. The irradiated samples show both residual stress variance among individual nanoscale subcascades and average residual hydrostatic and shear stresses at larger scales. This work supplements the hydrostatic residual stresses with residual orthorhombic shear and monoclinic shear stresses ($\sigma_O$, $\sigma_M$). As illustrated in Fig. 1 (right panel), a hydrostatic residual stress $\sigma_H$ (green line in the figure) may be sufficient to characterize the residual stress state in the special case of high-symmetry structures and isotropic radiation (homogeneous irradiation from all spherical directions). More general materials/radiation conditions, however, require the inclusion of shear tensor components (red regions in the figure). In the following, we document several scenarios that demonstrate the existence and magnitude of residual shear stresses after irradiation, including fixed-momentum PKA irradiation, low-symmetry compound with isotropic radiation, and high symmetry metals under pre-stress $\boldsymbol{\sigma}_{\text{pre}}$.

**3.1. Residual shear stresses under fixed PKA momentum**

As illustrated in Fig. 1, when an incident energetic particle (ion or electron) collides with an atom on a lattice site, it produces kinetic primary knock-on atoms (PKAs) and forms various radiation "subcascades" of a total radiation cascade, varying from each other chronologically and spatially. Since the ultrahigh resolution (~Å) of focused electron beams makes it possible to precisely control the post-collision momentum distribution of PKAs (3), we first consider fixed-momentum PKA irradiation, as demonstrated previously by Stoller et al. (39).

Our simulations show that radiation-induced shear stress exists and plays a critical role as Eshelby transformation. Here we reveal the existence of both residual shear stress fluctuation among single subcascade and average residual shear stresses across the entire irradiated zone, as illustrated in the left panel of Fig. 1. We divide the overall residual stress $\boldsymbol{\sigma}_{\text{Re}}$ into three tensor components: a hydrostatic stress $\boldsymbol{\sigma}_{\text{H}}$, an orthorhombic shear $\boldsymbol{\sigma}_{\text{O}}$, and a monoclinic shear $\boldsymbol{\sigma}_{\text{M}}$ (Fig. 1, left panel).

$$\boldsymbol{\sigma}_{\text{Re}} = \boldsymbol{\sigma}_{\text{H}} + \boldsymbol{\sigma}_{\text{O}} + \boldsymbol{\sigma}_{\text{M}} \tag{1}$$

To quantify the magnitude of the tensors $\boldsymbol{\sigma}_{\text{H}}$, $\boldsymbol{\sigma}_{\text{O}}$, and $\boldsymbol{\sigma}_{\text{M}}$, we convert them into scalars as shown in Supplementary Note 2. The combination of the orthorhombic shear stress $\boldsymbol{\sigma}_O$ and the monoclinic shear stress $\boldsymbol{\sigma}_M$ determines the magnitude of the shear stress, which can be represented by the second-order rotational invariant of deviatoric stress $II_\sigma$ or equivalently the von Mises stress $\sigma_{\text{von Mises}}$. All the stresses and axes are described based on the natural cubic coordinates of materials in the following sections. Here we decompose the total shear stress into $\boldsymbol{\sigma}_O$ and $\boldsymbol{\sigma}_M$ under the material natural coordinates, considering that the studied materials are all orthorhombic/tetragonal/cubic crystalline structures and the corresponding shear behaviors are different for $\boldsymbol{\sigma}_O$ and $\boldsymbol{\sigma}_M$ during radiation (discussed in the following contents). The evolution of $\boldsymbol{\sigma}_O$ and $\boldsymbol{\sigma}_M$ under radiation shows a strong correlation with material intrinsic structures. The void/cavity induced atom transfer has been shown as the main factor inducing volumetric expansion (27), whereas in this study, no high-radiation-dosed and biased long-timescale defect evolution is involved, therefore we expect the effects of residual stresses on volumetric swelling is relatively small.

To first study the residual stress as a result of PKA-initiated irradiation, we keep the initial momentum and positions of the PKAs fixed and vary the unirradiated equilibrium configuration in each trial (see Methods section for details). This procedure results in a distribution of residual stress. The average hydrostatic component ($\sigma_H V$) of the Kanzaki force dipole tensor is similar to the values reported by Dudarev et al. (40) and Garg et al. (14). Fig. 2 (also Fig. S2-S4) shows the probability distributions of residual shear stresses in BCC-iron and tetragonal zircon (ZrSiO$_4$) single crystal, respectively. Since the behaviors of three monoclinic shears are similar, we utilize $\sigma_{xy}$ as a representative. It is evident from Fig. 2 that the residual shear stress distributions show multiple peaks for low incident energies and single peaks for high incident energies. Such differences in the residual shear stress distributions can be attributed to the type and number of defects generated at different levels of PKA energy.

Low-energy irradiation often creates isolated defects of specific types, thus leading to distinct residual stress fields in individual cascade events. By contrast, high-energy irradiation simultaneously generates multiple defects of various types and creates a smeared-out average residual stress field. The relative absolute ratio between average $\sigma_{xy}$ and $\sigma_H$ ranges from 2.6% to 14.9%. The typical defect configurations causing these hydrostatic expansions and shears are shown in Fig. 2 as insets. In iron, the <111> dumbbells are the main sources of shear stress (41, 42) and the preference for specific orientations is the reason for net residual shear stress. Four typical orientations of <111> dumbbells and their corresponding shear characteristics are presented in Table S3.

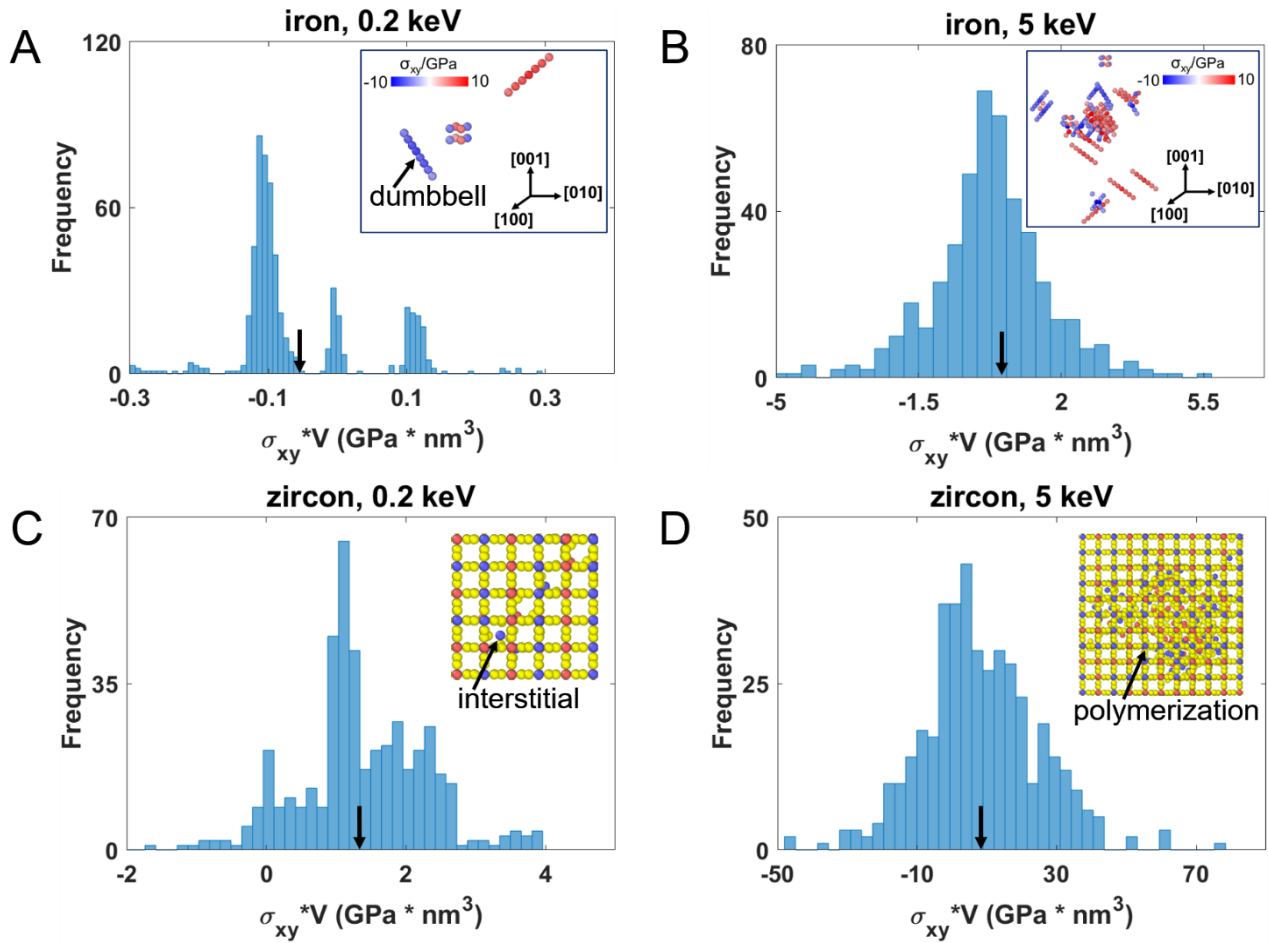

**Figure 2.** Distributions of radiation-induced residual shear stresses. (A and B) Distribution of the elastic dipole component ($\sigma_{xy}V$) in BCC iron. An incident energy of 0.2 keV in (A) only generates isolated defects (one or two <111> dumbbells as shown in the inset) in each trial. This results in the discrete distribution of ($\sigma_{xy}V$) with three

peaks. The incident energy in (B) is 5 keV which causes the interaction between defects, as shown by the inset. The multiple combinations of defects generate a continual distribution of ($\sigma_{xy}V$). (C and D) Distribution of the elastic dipole component ($\sigma_{xy}V$) in zircon. Point defects (inset of (C)) resulting from 0.2 keV radiation in (C) also correspond to a discrete distribution like iron, while amorphization (inset of (D)) and continual distribution appear in 5 keV radiation in zircon (D). The direction of all incident momentum vectors is fixed to be [122] and each case has been carried out with 500 trials. The average values of shear components have been denoted by the black arrows, corresponding to the values of -0.058, 0.26, 1.57, and 8.28 in (A), (B), (C), and (D), respectively. In the insets of (A) and (B), atoms with von Mises stresses lower than 1 GPa are hidden.

Since $ZrSiO_4$ crystal is composed of three elements and its structure is not cubic, the causes of the irradiation shear stress are more complex. When the incident energy is low, the residual shear stress is mainly due to various types of point defects (interstitials, vacancies, and antisites), while defect amorphization occurs and causes shear stress as a result of high-energy irradiation (Fig. 2C and 2D, inset). The defect-amorphized phase, consisting of connected $SiO_n$ polyhedra, is observed in the damaged region, stabilizing the defect type and stress evolution within the region. Fig. S5 specifically discusses MD snapshots of different types of point defects and the corresponding residual shear stresses in zircon. Fig. S6 shows the defect-amorphized phase and the formation of bridging oxygens after high-energy irradiation.

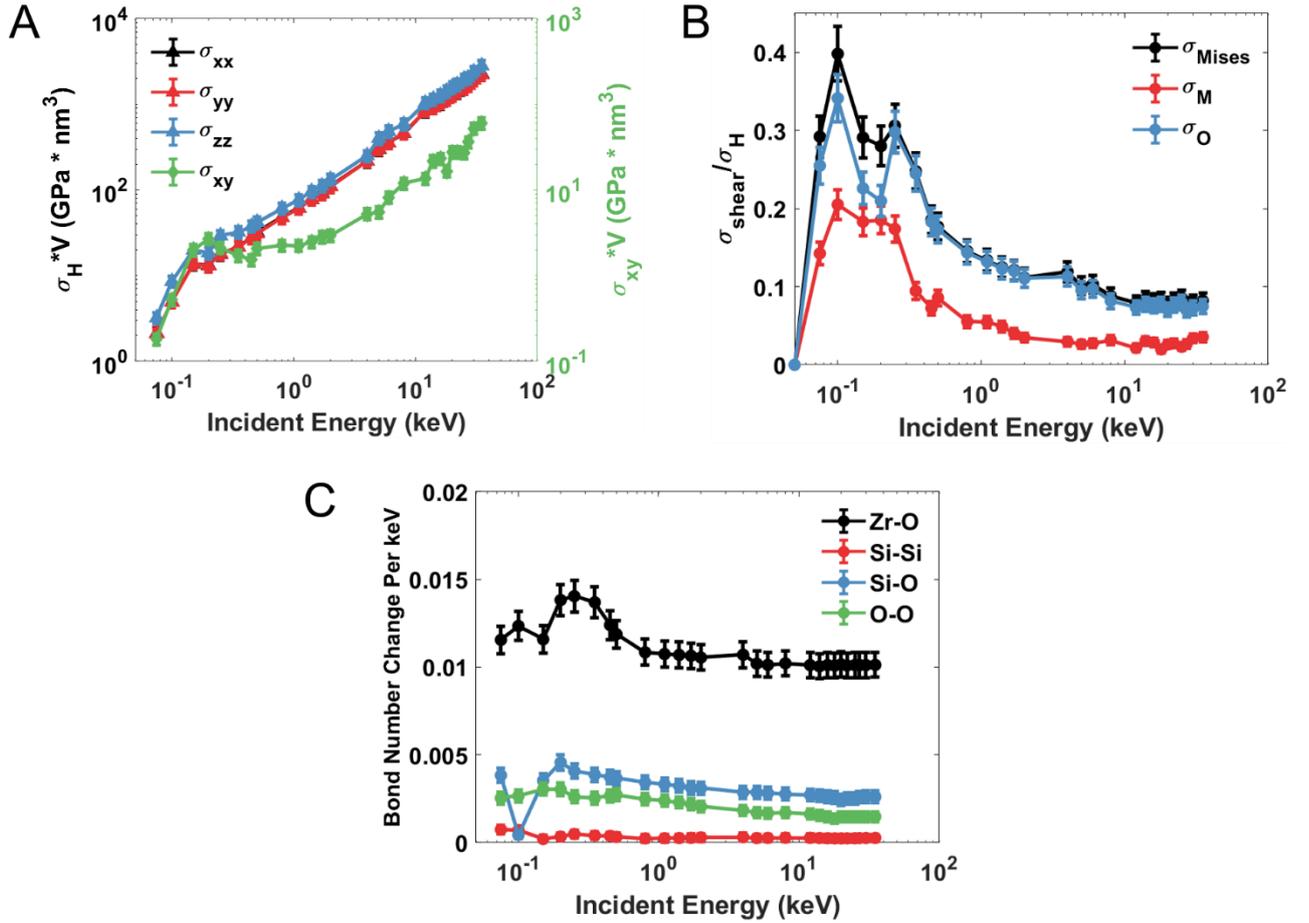

**Figure 3.** Characteristics of average residual shear stresses in zircon. (A) Energy dependence for the elastic dipole tensor of damage defects. The three normal stresses remain negative in all cases, while here, we invert their sign for consistency. Here $(\sigma_{zz}V)$ is always higher than $(\sigma_{xx}V)$ and $(\sigma_{yy}V)$, indicating the existence of residual orthorhombic shear. This phenomenon is attributed to the tetragonal anisotropy of the atomic structure. The component $(\sigma_{xy}V)$ is plotted as the representation of three monoclinic shear stresses. (B) Evolution of shear to hydrostatic stress ratio with energy. The ratio arises initially since low radiation energy shows a low probability of creating defects. The later declining behavior is a result of large scattering in high-energy radiation. The ratio comes to the same level under high energy radiation, as a result of damage fragmentation. (C) Defect microstructural evolution with energy. The average bond number changes in all cases are counted and divided by incident energy. The behaviors of initial fluctuation and later convergence correspond well to both stress evolution in (A) and ratio changes in (B). All the parameters in (A-C) are the average values extracted from 30-150 trials in each setup, respectively, until that two times standard deviations (error bars) are less than 10%.

To further explore the coupled effects of momentum directionality and magnitude on the average residual shear stress, we systematically tune the incident energy of the PKAs while fixing the direction to be [110], as shown in Fig. 3. In general, the relative magnitude of the residual shear stresses with respect to $\sigma_H$ first increases with increasing energy and then levels off to a certain steady level with even further energy increase, e.g., $\sigma_{\text{von Mises}}/\sigma_H$ drops from ~40% at its peak to about 9% at the tail. This kind of behavior can be rationalized by the large scattering effect in high-energy irradiation and is consistent with the microstructural evolution, as shown in Fig. 3C. Meanwhile, the correlated convergence of residual stresses in Fig. 3A and 3B and defect types in Fig. 3C further strengthens our claim that the preference for certain defect types causes residual shear with a *non-zero mean*. Under the stimulus of a monodisperse ion momentum in a low-symmetry direction, average residual shear stresses exist in a wide range of incident energies for different materials.

### 3.2. Residual shear stress in low-symmetry materials with isotropic radiation

Next, we show that the intrinsic structural asymmetry is another important factor that contributes to the formation of shear stresses upon irradiation. In the above section, we show that the lower symmetry of $ZrSiO_4$ crystal results in radiation-induced residual shear stress, see Fig. 3A, where the irradiation-induced $\sigma_{zz}$ is always higher than $\sigma_{xx}$ and $\sigma_{yy}$, suggesting the existence of a shear component $\sigma_T$. However, it needs further clarification whether the observed residual shear stress is mainly due to the momentum stimulus or the crystal structure anisotropy, i.e., if the atomic disparity between one crystal axis (*z*-direction) and the other two (*x*- and *y*-directions), Fig. 4A, is an independent factor contributing to the residual shear stress, even in the absence of momentum monodispersity. To that end, we use relatively symmetrical iron and asymmetrical zircon for a comparison. To demonstrate the effects of structural asymmetry on residual shear stresses in the absence of momentum directionality, we considered 100 non-parallel incident directions of PKAs (Fig. S7) and calculated the average residual stress.

As noted in Table 1, the shear stress is very small compared to the hydrostatic stress for high-symmetry iron, but that is not the case for zircon. Fig. 4B shows $\sigma_O$ and $\sigma_O/\sigma_H$ as a function of the incident energy for random incident directions. The $\sigma_O$ stress component in zircon increases with energy and is a significant fraction of the hydrostatic stress. At high incident energy, the shear stress increases in proportion to the hydrostatic stress and is approximately 10% of the latter. By contrast, the residual stress in iron is very nearly hydrostatic over the entire range of incident energy. As revealed by previous studies, the anisotropic deformation of low-symmetry materials during radiation (e.g., radiation growth) varies significantly under different radiation doses (43-45). The above results then hint that the material's non-hydrostatic deformation will be shaped by both internal crystal low-symmetry and external momentum monodispersity.

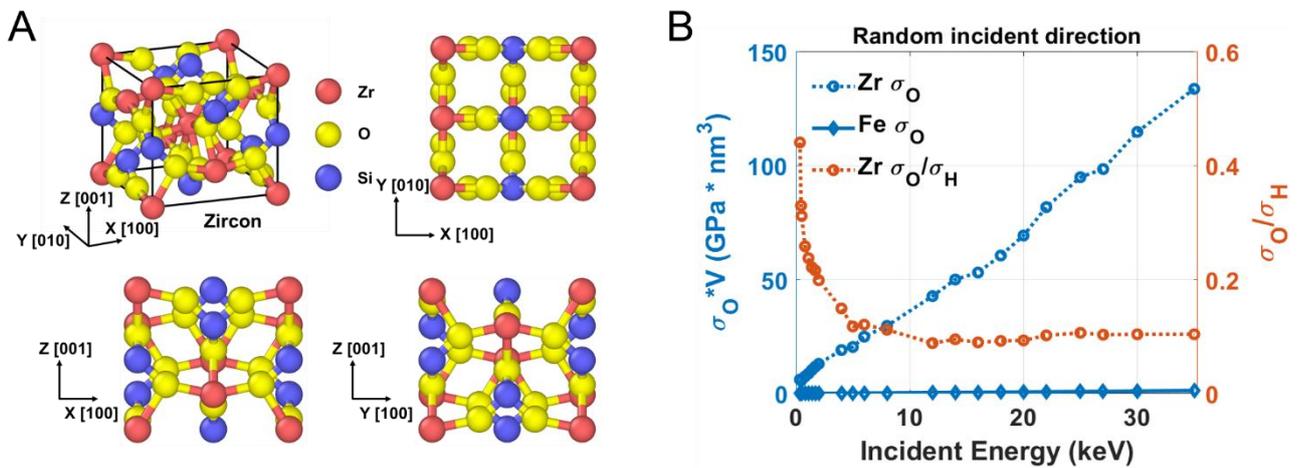

**Figure 4.** Structural asymmetry induced average residual shear stresses. (A) Atomic structure of zircon and its projections on *xy*, *xz*, and *yz* planes. The projection on *xy* plane is different from *xz* and *yz* planes, while the later two projections own mirror symmetry. (B) Energy dependence for the ($\sigma_O V$) and ($\sigma_O/\sigma_H$) of damage defects. All the parameters in (B) are the average values extracted from 100 cases with spherical incident directions, respectively.

**Table 1.** Average residual stresses with 2 keV energy stimulus and no momentum directionality. Both orthorhombic shear and monoclinic shear are small for iron, while the orthorhombic shear in zircon is non-negligible compared to hydrostatic stress.

| Materials | $\sigma_H V$ (GPa $* nm^3$) | $\sigma_O V$ | $\sigma_M V$ | $\sigma_{\text{von Mises}}$ to $|\sigma_H|$ ratio |
|---|---|---|---|---|
| iron (high symmetry) | -2.412 | 0.0066 | 0.0198 | 0.0086 |
| zircon (low symmetry) | -73.5386 | 14.6341 | 0.1164 | 0.1990 |

Therefore, to describe the stress response in a more general material and irradiation space, we need to take into account both the hydrostatic component and shear components:

$$\boldsymbol{\sigma}_{\text{Re}} = \boldsymbol{\sigma}_H + \boldsymbol{\sigma}_{\text{von Mises}}(\boldsymbol{p}, \boldsymbol{A}) \tag{2}$$

where $\boldsymbol{\sigma}_H$ is the conventional hydrostatic term, $\boldsymbol{\sigma}_{\text{von Mises}}(\boldsymbol{p}, \boldsymbol{A})$ is the residual shear stress due to momentum monodispersity $\boldsymbol{p}$ and material asymmetry $\boldsymbol{A}$. For high-symmetry materials under isotropic ion flux, the shear stresses approach zero and Eqn. (2) represents the conventional description of the irradiation response. Both high crystal symmetry ($\boldsymbol{A} \to \boldsymbol{0}$) and isotropic radiation flux ($\boldsymbol{p} \to \boldsymbol{0}$) can be considered as special cases; practical applications usually involve more general materials and conditions of irradiation.

**3.3. Residual shear stress in high-symmetry materials with nonhydrostatic pre-strain**

### 3.3.1. Oriented cluster induced residual shear stress in iron

Following the above demonstration of the effect of material asymmetry on radiation-induced actuation stresses, we now show that, even for materials with relatively high symmetry, externally applied loads can create sufficient asymmetry in the Eshelby transformation strain $\epsilon_{\text{transform}}$ to develop upon irradiation. This is actually just the single-crystal-level irradiation creep effect $\Delta\epsilon_{\text{transform}}(\boldsymbol{\sigma} > 0, K(\boldsymbol{p}))$ where $K(\boldsymbol{p})$ is the fluence of momentum $\boldsymbol{p}$. It is on top of the aforementioned irradiation swelling effect $\epsilon_{\text{transform}}(\boldsymbol{\sigma} = 0, K(\boldsymbol{p}))$, that can cause a transformation strain even when the pre-stress $\boldsymbol{\sigma} = 0$. We impose a pre-stress $\boldsymbol{\sigma}$ on iron and irradiate it with kinetic ions in random directions multiple times. Fig. 5A shows the evolution of the residual $\sigma_{xy}$ during continual self-ion irradiation, in which an elastically pre-strained state is acquired under pre-stress $\bar{\sigma}_{xy} = 0.6$ GPa and $\bar{\sigma}_{xx} = 0.6$ GPa. Under the coupled stimuli of irradiation and external stress, both $\bar{\sigma}_{xy}$ and $\bar{\sigma}_O$ drop to lower levels with increasing cascade events (as shown in Fig. S8A and S8B), indicating the partial transformation of the applied elastic pre-strain into plastic strain. The associated stress relaxation is referred to as irradiation creep and occurs when defects are generated to plastically accommodate the prescribed stresses. Accordingly, these defects generate residual strains (transformation strains) after the removal of the applied pre-stress. The corresponding residual stress $\sigma_{\text{Re}}$ in Fig. 5A is then obtained by subtracting the initial stress state $\bar{\sigma}_{\text{initial}}$ from the current state $\bar{\sigma}_{\text{current}}$:

$$\boldsymbol{\sigma}_{\text{Re}} = \bar{\boldsymbol{\sigma}}_{\text{current}} - \bar{\boldsymbol{\sigma}}_{\text{initial}} \tag{3}$$

We note that in our case, the residual stress field does not entirely offset the pre-strains and saturates when reaching a certain radiation dose. Similar to magnetic or electric polarization, we define a 'saturation point' at which the induced shear stresses start to level off, as marked in Fig. 5A. Compared with the curves of 20 keV irradiation, irradiation with 5 keV ions saturates with more cascade events and the released stress is much larger. The ratios of shear stress to hydrostatic stress are shown in Fig. 5B for both energy levels (also see Fig. S8C). There is a large difference in the $\sigma_{xy}/\sigma_H$ ratio depending on the energy level, whereas the $\sigma_O/\sigma_H$ evolution is quite close. These results clearly suggest that even high-symmetry materials, when subjected to non-hydrostatic pre-strains, can show significant additional shear transformation strains after radiation.

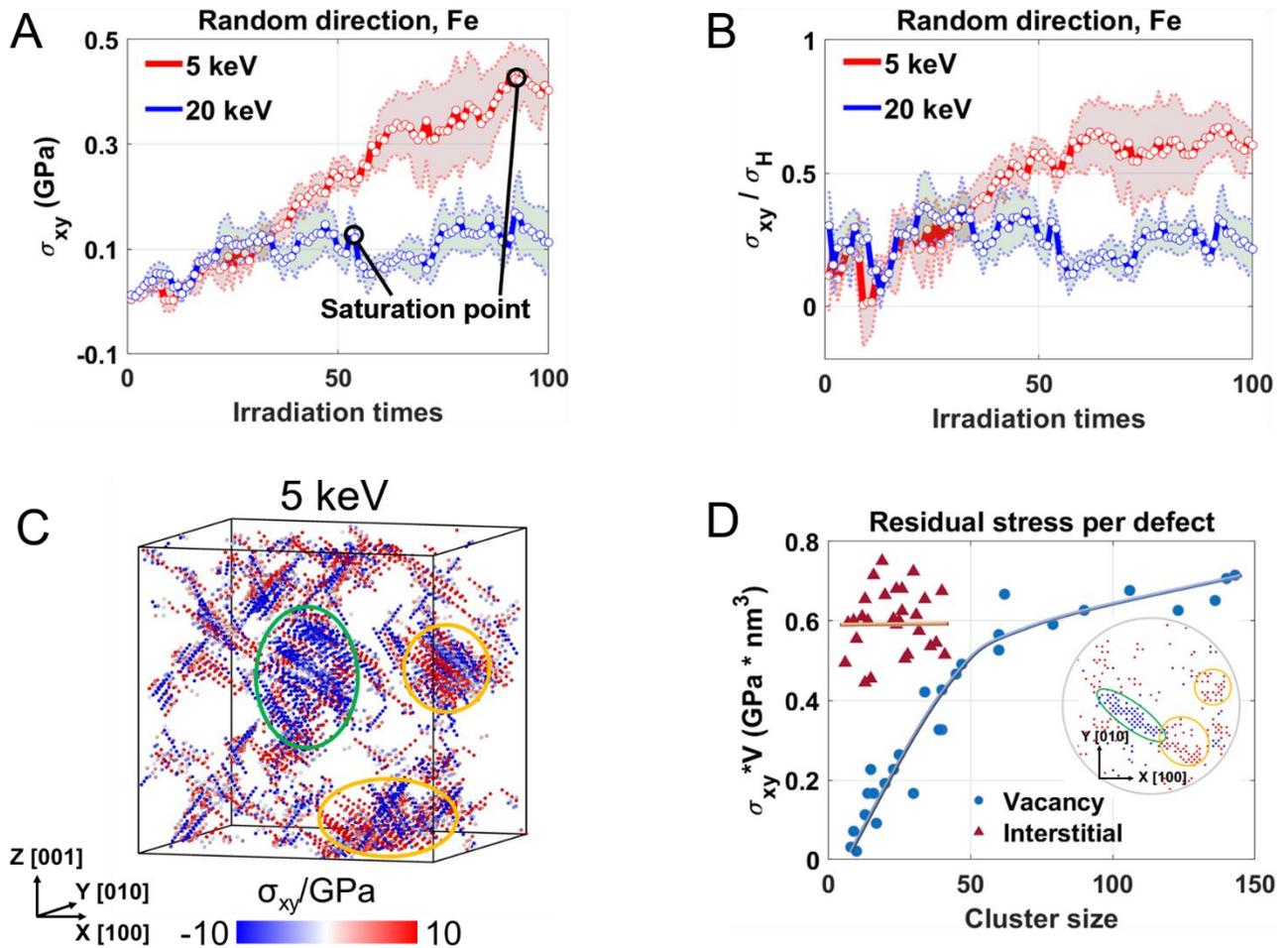

**Figure 5.** Tunable material asymmetry and residual stresses with pre-strain on iron. (A and B) Evolution of residual shear stresses and corresponding shear to hydro ratios under continual irradiation. In each case, red and blue curves represent the average shear stresses of 5 trials with an incident energy of 5 and 20 keV, respectively. The light-colored region demonstrates the error bar, whose boundary is equal to 2 times of the standard deviation. Two typical saturation points are circled in (A), marking the stabilization of stresses. (C) Color mapping of atomic stress ($\sigma_{xy}$) distribution in iron after 100 times irradiation of 5 keV self-ion. Atoms whose von Mises stresses lower than 1 GPa

are hidden. The major vacancy and interstitial clusters are denoted by green and orange circles, respectively. (D) Cluster size-dependent residual shear stress ($\sigma_{xy}V$) per interstitial/vacancy. Each ($\sigma_{xy}V$) is extracted by adding the shear component ($\sigma_{xy}$) multiplying ($V$) of the whole cluster region, and then normalized by cluster size. This demonstrates the average magnitude of residual shear stress induced by single defect in a specific sized cluster. Each data point represents one counted cluster located in post-irradiated sample. The inset shows the project of defect coordinates on xy plane, corresponding to the snapshot of (C). The green and orange circles represent vacancy (blue atoms) and interstitial (red atoms) clusters, respectively.

To reveal the underlying physical mechanisms, we first analyze the distribution of interstitials (red atoms) and vacancies (blue atoms) in a pre-stressed iron sample after irradiation. As shown in Fig. S8D, the sample shows significant defect aggregation. Fig. 5C shows the atomic configuration of the defect clusters in the sample. The atoms are colored depending on the local value of $\sigma_{xy}$; vacancy and interstitial clusters are marked by green and orange circles, respectively. In our earlier discussion of iron samples irradiated a single time (Fig. 2), we attributed the shear stress to interstitial-induced anisotropic volume expansion, i.e., oriented dumbbells, and conversely considered the volumetric change induced by a single vacancy to be nearly isotropic. However, in Fig. 5C, vacancy clusters counterintuitively show huge shear transformation stress. In other words, residual shear stresses under continuous irradiation can be attributed to anisotropic shape changes associated with 1) preferentially oriented dumbbells caused by point or aggregated interstitials and with 2) aggregated vacancies.

### 3.3.2. Cluster size-dependent Eshelby shear transformation

We further look into how vacancy clusters produce shear transformation strain in pre-stressed samples. We note that vacancy clusters that form in pre-stressed samples often have a disk-like shape (see Fig. 5D inset for the projection of defect snapshot in Fig. 5C onto the *xy* plane), in contrast to clusters in unstrained samples, which have a nearly spherical shape, see Fig. 6B. We conjecture that the pre-stress promotes the formation of vacancy clusters in non-spherical shapes, which in turn induces a residual or transformation shear strain, as one would expect based on an Eshelby analysis. We verify this hypothesis by artificially creating a vacancy cluster in a pre-stressed perfect sample and equilibrating the whole system. As shown in Fig. 6D, the system forms a stable oblate vacancy cluster. Thus, under the conditions of continuous irradiation and external pre-strain, materials tend to form defect clusters whose stress field is in alignment with the pre-strain.

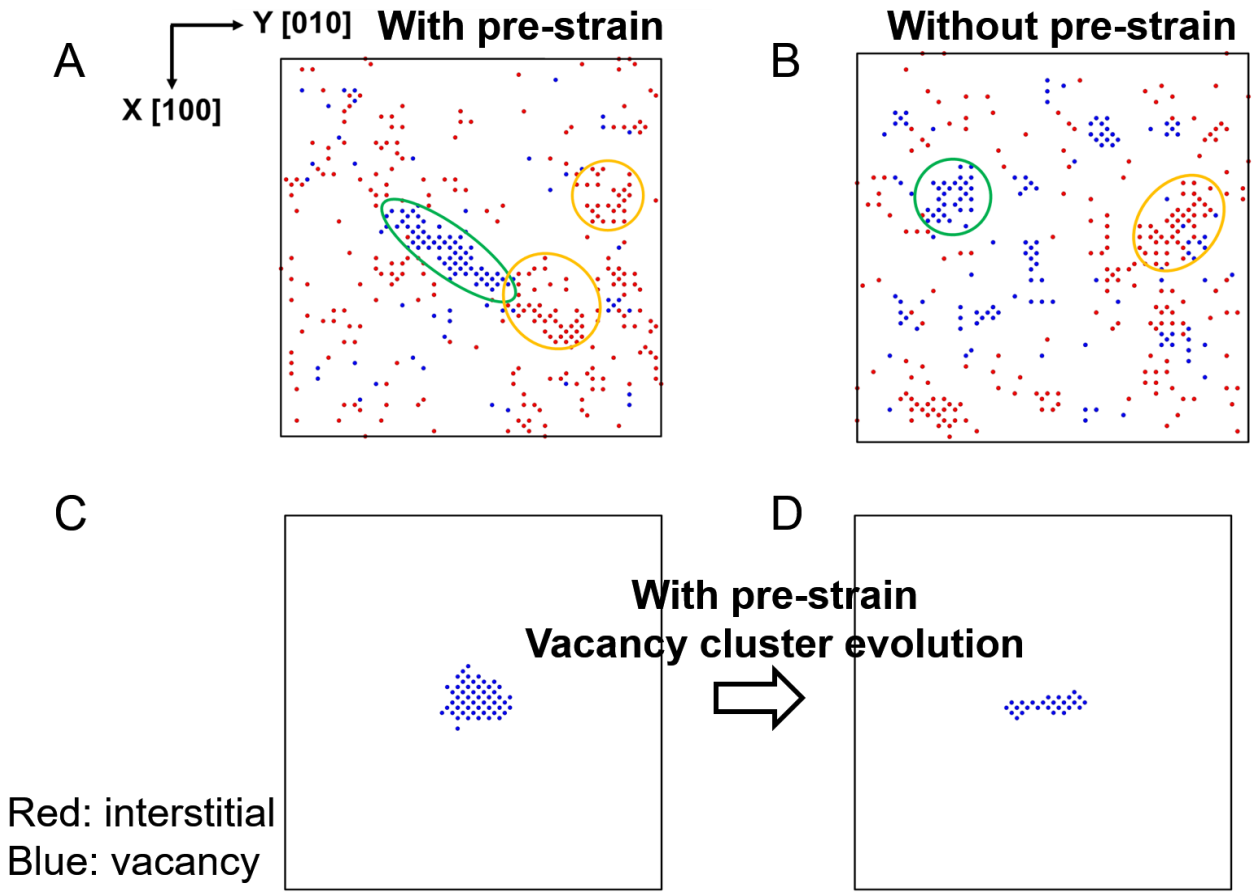

**Figure 6.** Morphology of defect clusters under anisotropic pre-strain. (A and B) Coordinate projections of defects on xy plane for post-irradiated systems. In the pre-strained system (A), the global shear stress transforms the vacancy cluster (green circled) into a disk-like shape, thus inducing large residual shear response. In the strain-free system (B), the cluster morphology of vacancies still maintains the near-spherical shape. The interstitial clusters in both systems do not experience significant shape transformation. (C and D) Verification of shear strain's role in morphology evolution of clusters. In a pre-strained sample, one vacancy cluster is manually created with a near-spherical shape in (C). This vacancy cluster finally reaches a disk-like shape (D) under global shear's impact.

Single vacancies result in nearly isotropic residual stress fields, whereas aggregated vacancies have anisotropic stress fields that can interact with externally applied shear strains. Hence, inclusions with varied shapes generate different residual stress field, which in this case strongly correlates with cluster sizes. Fig. 5D shows the average residual shear component per interstitial/vacancy, for different-sized clusters. To quantify this, the shear dipole ($\sigma_{\text{von Mises}}V$) is normalized:

$$\sigma_{\text{von Mises}}^{\text{mono}}V = \frac{\sum_{\text{cluster}} \sigma_{\text{von Mises}}V}{N} \tag{4}$$

Here, $\sigma_{\text{von Mises}}^{\text{mono}}V$ refers to the shear dipole induced by a single defect and $N$ is the total number of defects in each cluster. Fig. 5D shows $\sigma_{\text{von Mises}}^{\text{mono}}V$ as a function of cluster size. It is evident that the aggregation of vacancies in non-spherical clusters increases the shear character of the stress induced by vacancies, while the stress contribution from interstitials is largely independent of the cluster size.

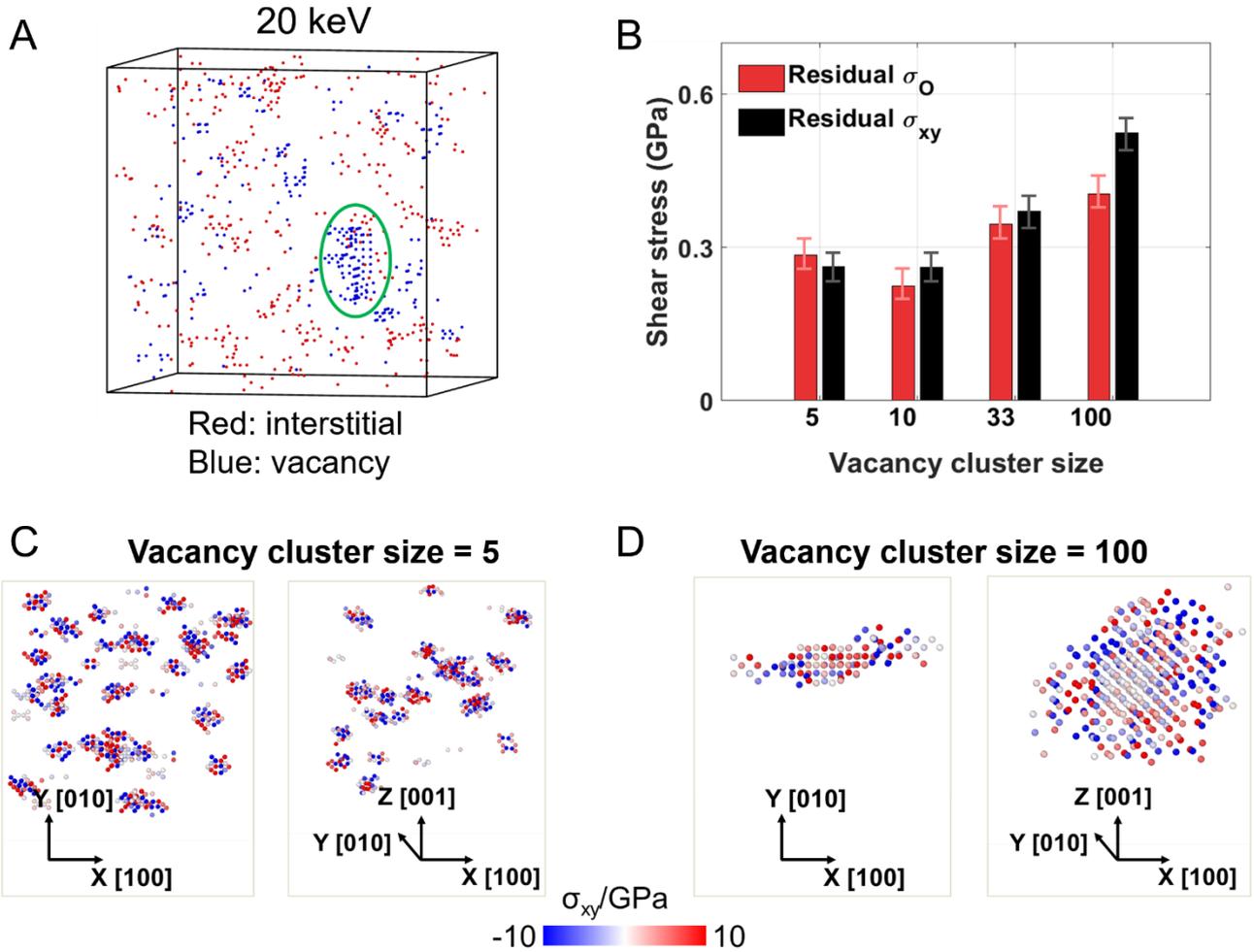

**Figure 7.** Dependence of residual shear stress on defect cluster size. (A) MD snapshot of defect distribution in 20 keV irradiated system. The size of the largest vacancy cluster (green circled) is smaller than that in 5 keV irradiated system in Fig. S8D. (B) Released shear stress vs. vacancy cluster size. The larger the cluster, the more significant shape change and more residual shear responses are induced. (C and D) Size-dependent vacancy cluster morphology and residual stress. In pre-strained samples, vacancy clusters are artificially introduced with the average cluster sizes ranging from 5 to 100, and the total vacancy number is held to be 100. The vacancy cluster aggregates into random shapes when the size is 5 (C). However, the large cluster with 100 vacancies will deform into a disk-like shape under the impacts of shear strain (D). Here we only show the atoms whose von Mises stresses are higher than 1 GPa.

Fig. 5(A and B) and Fig. S8(B and C) show the lower residual shear stresses in 20 keV irradiation than the 5 keV case. In contrast to the 5 keV case (Fig. S8D), irradiation with 20 keV energy (Fig. 7A) creates defects that are more dispersively distributed and the corresponding cluster size is smaller. A possible explanation for this cluster size trend is that the annealing effect arises and hinders defect aggregation in high incident energy (20 keV) cases.

Thus, there should be a strong dependence of residual shear stress with the created cluster size. To justify this hypothesis, we manually introduce vacancy clusters of different sizes into the perfect pre-strained sample. The average sizes range from 5 to 100 in different samples, while the total vacancy number is controlled to be 100 in all cases. Comparing the equilibrated cluster morphologies in Fig. 7C and 7D, the larger cluster tends to extend into the disk-like shape and the smaller cluster is still holding a near-spherical shape. As for the released shear stresses in Fig. 7B, there is an apparent escalation when the average cluster size is greater than 10. Therefore, under the stimulus of external stress, small vacancy cluster still keeps nearly isotropic volumetric change, whereas the large aggregation deforms and responds anisotropically to the outer stress field. The above equilibrium tests suggest that defect cluster size inversely depends on the magnitude of incident energy, while a larger cluster size tends to form higher internal residual stress field. This conclusion is highly consistent with the size-dependent shear in irradiation in Fig. 5D, and explains how incident energy affects residual stress fields.

### 3.3.3. Non-creep components of radiation-induced shear stress in copper

According to the discussion on iron irradiation, irradiation on pre-strained sample will induce great irradiation-creep behavior and arouse significant residual stress. However, we also find that in some cases non-creep residual stress components will be aroused. For instance, Fig. 8A presents the stress evolution of the pre-strained copper during continual irradiation. Under a tensile stress of $\sigma_{xx}$ = 1.5 GPa, the sample not only experiences irradiation creep to release $\sigma_{xx}$, but also generates a significant residual $\sigma_{xz}$ (brown colored in Fig. 8A). Thus, external tensorial stimulation can induce a non-parallel tensorial response. Fig. 8(B-D) illuminates the defect origin of this interesting residual stress response. There are two MD snapshots in Fig. 8B corresponding to the escalation process of ($\sigma_{xz}$). Comparison of moments I and II reveals the expansion of dislocation loops during irradiation, leading to a huge positive ($\sigma_{xz}$) region (red-colored atoms in Fig. 8C). It turns out that the gradual formation and expansion of prismatic dislocation loops lead to a $\boldsymbol{\sigma}_{Re}$ non-parallel to the initial $\boldsymbol{\sigma}_{pre}$. Thus, both the magnitude and orientation of residual stress response should be well studied in various materials under distinctive stimuli.

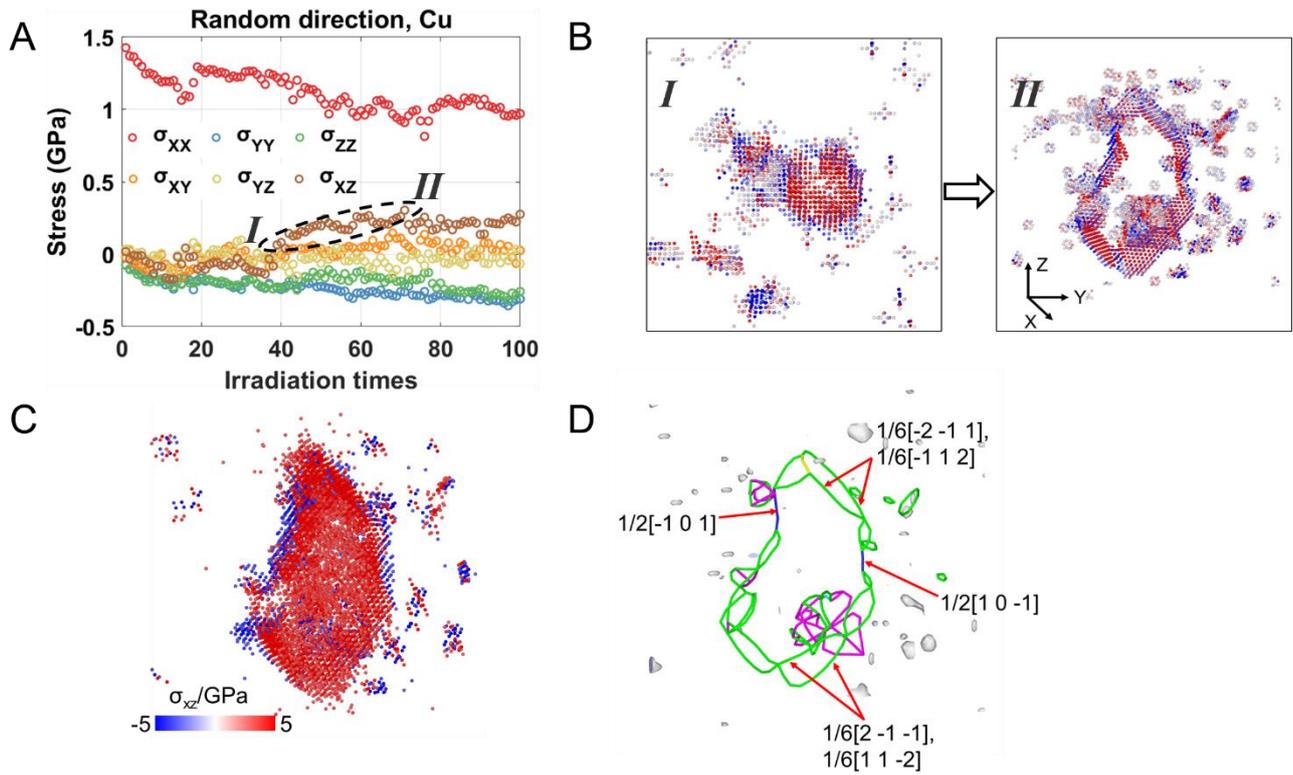

**Figure 8.** Residual shear response for copper under irradiation. The copper sample is pre-strained in the *x* direction with ($\sigma_{xx}$) equal to 1.5 GPa. (A) Stress evolution with irradiation times. Apart from the common stress relaxation caused by irradiation creep for ($\sigma_{xx}$), ($\sigma_{xz}$) (brown points) also rises greatly during continual irradiation. Thus, external tensorial stimulation can induce a non-parallel tensorial response. (B-D) Defect origins of residual shear stress ($\sigma_{xz}$). There are two MD snapshots in (B) corresponding to the escalation process of ($\sigma_{xz}$). Comparison of moments *I* and *II* reveals the expansion of dislocation loops during irradiation, leading to a huge positive ($\sigma_{xz}$) region (red-colored atoms in (C)). Defect structures in (B) and shear stress state in (C) are visualized by hiding atoms whose absolute hydrostatic stresses and von Mises stresses are respectively lower than 1 GPa. As shown in (D), this dislocation loop contains two prismatic dislocations whose burgers vectors are opposite. The dislocations are identified by dislocation extraction algorithm (DXA) (38). Green lines are Shockley partial dislocations, pink lines are the Stair-rod dislocations, yellow lines are Hirth dislocations, and blue lines are perfect dislocations.

Apart from pre-stress, pre-existing defects (grain boundaries (GB), dislocations, second phase particles, voids etc.) can also change lattice constants in an anisotropic way, thus introducing structural anisotropy into the material. These pre-existing sinks or strains vary spatially and interact with radiation defects, generating even more complicated tensor responses. We can predict a specific tensor relation through fitting for further tuning:

$$\sigma_{\text{Re}}^{ij} = C_{(1)}^{ijk} p_{\text{k}} + C_{(2)}^{ijmn} p_{\text{m}} p_{\text{n}} + \cdots \tag{5}$$

$\boldsymbol{C}_{(1)}$ and $\boldsymbol{C}_{(2)}$ are respectively third-order and fourth-order tensors, indicating the momentum and energy characteristics. $p$ values are momentum stimuli. The coupling of $p_{\text{m}}$ and $p_{\text{n}}$ includes the energy portion. Higher-order tensors are reasonable and indispensable since coupled effects of momentum, energy, material structures, pre-stress states, and phase transitions are exceedingly complex.

### 3.4. Applications of average residual shear stress for irradiation shaping

Material modification and defect creation can be precisely controlled if the response of a material to a FIB is well understood. Here, we suggest a method for steering the dynamical material modification process towards the desired nano-morphology by taking advantage of the tensor response under monodisperse-momentum irradiation. We believe this method can eventually lead to "3D Printing of Eshelby Transformations" technology in small material components, as FIB has nm-resolution in *x*, *y*, and the depth of implantation *z* (Bragg peak) can be controlled by the ion energy (6). Typically with heavy ions, *z* is on the order of 1μm; but with light ions such as proton, *z* can be sharply defined up to hundreds of μms.

To demonstrate the role of the average residual shear induced by ion irradiation, we conduct further self-ion irradiation simulations. In this section, the initial kinetic particles are not selected from the original atoms (treated as PKAs) on lattice sites as in the previous sections, but are introduced externally into the sample with changed initial positions around the lattice (Fig. S9). Hereafter, we discuss the results obtained for zircon, in which the coupling of lattice asymmetry and momentum are representative.

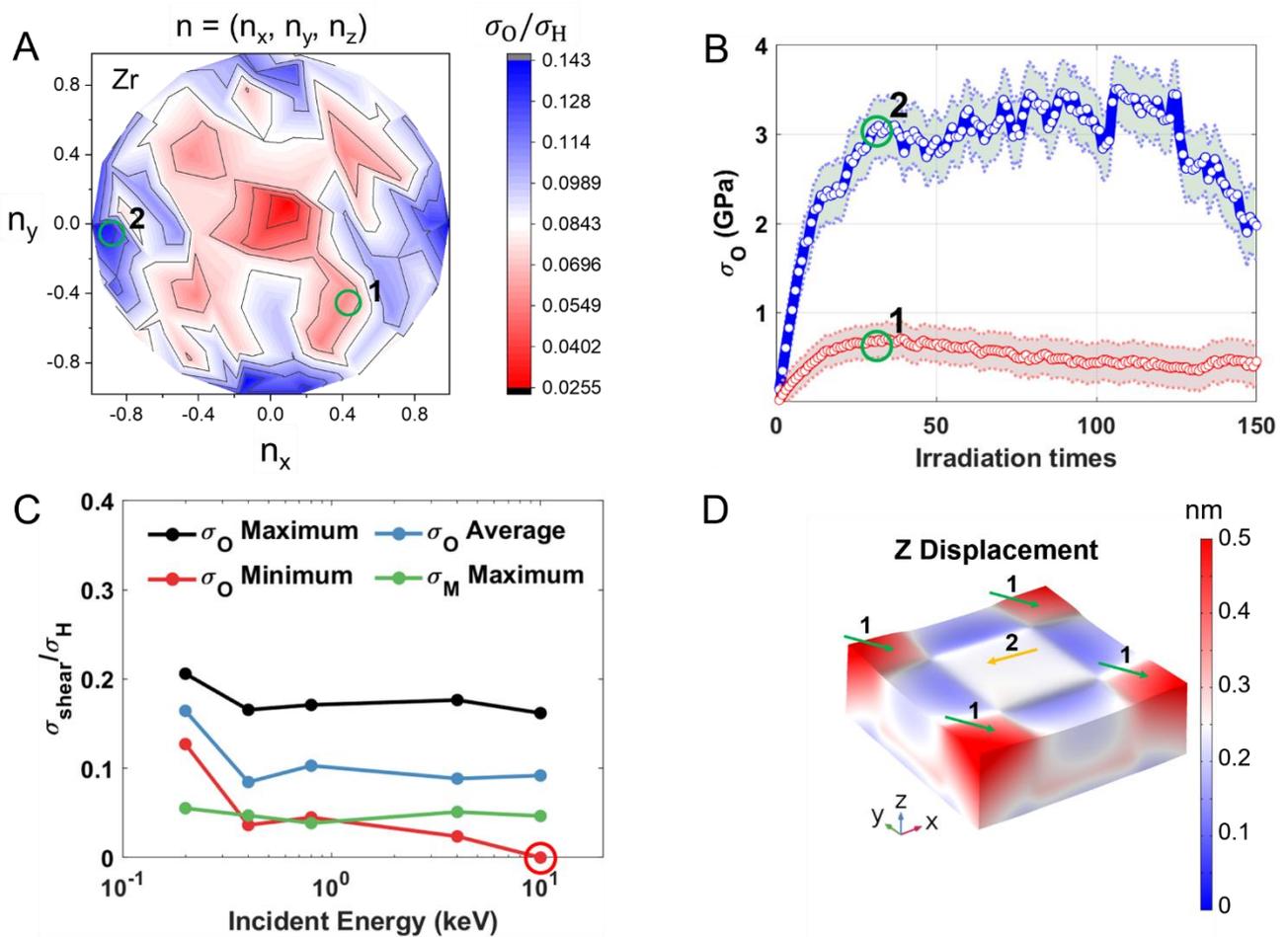

**Figure 9.** Application of self-ion radiation-induced residual shear in defect engineering. (A) The ratio $\sigma_O/\sigma_H$ versus incident orientations under 0.4 keV irradiation. Here we plot 2D polarization on the *xy* plane, in which the polar angle is consistent with the incident 100 spherical directions. The *z* component $n_z$ can be calculated from $n_x$ and $n_y$. The polar plot generally owns mirror symmetry on three planes due to the crystallographic symmetry of zircon. (B) The evolution of $\sigma_O$ with irradiation times under two typical directions as denoted in (A). (C) Dependence of $\sigma_O$ Maximum, $\sigma_O$ Minimum, $\sigma_O$ Average, and $\sigma_M$ Maximum with energy among 100 directions. (D) FEM model of defect engineering employing the stress states (green circled) acquired from two incident directions in (B).

To correlate the average shear behavior with the momentum vectors of the self-ions, 100 non-parallel spherical irradiation directions are implemented as described above. Polar plots of the ratios of shear to hydrostatic stress are shown in Fig. 9A and S10A, respectively, corresponding to $\sigma_O$ and $\sigma_{xy}$ with a monochromatic kinetic energy of 0.4 keV. Fig. 9A indicates a strong orthorhombic shear response along the *x* and *y*-axes and a nearly isotropic response along the diagonal directions and the *z*-axis. Fig. S10A denotes the significant monoclinic shear response (positive or negative) along the diagonal directions. The $\sigma_{xz}$ and $\sigma_{yz}$ polarizations in zircon are displayed in Fig. S10(B and C). <span style="color:red">The above results demonstrate a broad range of shear to hydrostatic ratios (2% to 15% of $\sigma_O/\sigma_H$, -7% to 7% of $\sigma_{xy}/\sigma_H$) resulting from different momentum vectors</span>; the distribution is diverse and complex yet reflects many characteristics of the underlying crystal structures.

Conceivably, combining the datasets generated from accurate simulations and experimental verification, we could apply machine learning to better understand and fit the tensor response displayed in Eqn. (5) in the future. Thus, the optimal momentum vector could be obtained so as to efficiently shape the nanomaterials by 3DPET. Fig. 9B illustrates an example for application in a nanofabrication process. Here, we extract two typical directions from the shear to hydrostatic distribution in Fig. 9A, with the most hydrostatic and shear responses denoted as directions 1 and 2. Ion doses are gradually increased by introducing multiple kinetic self-ions one at a time, which can be easily controlled with FIB. Fig. 9B shows the dependence of the whole orthorhombic shear on irradiation events (here, the total amount is 150). The $\sigma_O$ in both cases ascends proportionally with ion doses and then fluctuates or descends, indicating the accumulation of anisotropic response in the early stage and the later decline due to overlap of irradiation subcascades (32). The significant difference in $\sigma_O$ evolution between the two curves reveals the critical role of momentum stimulus. Predictably, further exploration of this "stress-strain"-like evolution curve is needed to first build up the benchmark for characterization, and then design functional materials with either ultrahigh anisotropic response or conversely robust anti-polarization property.

Further investigations of modulating anisotropic responses and effects of energy variation are presented in Fig. 9C (with the energy ranging from 0.2 keV to 10 keV). Each energy corresponds to 100 spherical directions with 100 trials carried for each setup to acquire the average values, from which 4 typical shear properties are extracted. The maximum ratios of $\sigma_O$ and $\sigma_M$ relative to $\sigma_H$ among these 100 cases level off at about 18% (black curve) and 6% (green curve), respectively. As for the minimum $\sigma_O$ to $\sigma_H$ ratio, the phenomenon that residual $\sigma_{zz}$ is always higher than $\sigma_{xx}$ and $\sigma_{yy}$ is no longer rigidly true and can be inverted if the momentum stimulus is strong enough, as shown by the red curve in Fig. 9C. Thus, we can well tune the range of $\sigma_O/\sigma_H$ and $\sigma_{xy}/\sigma_H$ by momentum variations.

Local modification of materials and defect implantation via focused ion beam at micro/nano scale has been applied to quantum optics (46) and circuits (47). Combining the anisotropic response with defect engineering, we can precisely produce targeted morphology and induce deep elastic strain in targeted regions by 3DPET. As visualized with FEM in Fig. 9D, here we implant two states of residual stress acquired from the irradiation process in Fig. 9B (marked by the green circles), and control their distribution with state 1 corresponding to the corners and state 2 corresponding to the main body. Though the volumetric expansions of these two states are close, heterogeneous deformations can help modify the sample into a designated morphology and induce high von Mises stress at the intermediate area. Another application could be the stress engineering in semiconductor manufacturing, where the artificial dislocations can be precisely implanted to induce a compressive stress state for better conductivity (48). Our research then supplements that the shear state can also be controlled. Moreover, former studies revealed that a mild tensile strain of 0.4% is sufficient to achieve a fundamental direct bandgap in $Ge_{0.94}Sn_{0.06}$ binary alloys (49). It has been observed that bandgap changes can be as large as 60 and 100 meV per 1% axial strain for [100] and [110] silicon nanowires, respectively (50, 51). These results open up new avenues to induce nanoscale mechanical loading with irradiation modification in quantum manufacturing and deep elastic strain engineering (13).

4. **Discussion**

One important benefit of using an ion beam to modify a material is that the radiation dose can be precisely adjusted and delivered with nanometer-scale precision. Here, we supplement these advantages of ion-beam fabrication by establishing that ion-beam irradiation not only generates hydrostatic residual stresses but also shear residual stresses with certain mean and variance, which can be judiciously tuned by the energy and direction of incident electrons or ions, intrinsic material asymmetry, and external loadings such as non-hydrostatic pre-strains or pre-existing defects (52, 53).

We have seen that a pre-stress is particularly effective in creating additional residual strain,

with a near 70% shear-to-hydrostatic ratio. If we call the Eshelby transformation induced at $\sigma_{\text{pre}}=0$ by a specific radiation stimulus "generalized radiation swelling", then the difference in the Eshelby transformation strain-volume between $\sigma_{\text{pre}} \neq 0$ and $\sigma_{\text{pre}}=0$ under the same radiation can be called "generalized radiation creep". Both "generalized radiation swelling" and "generalized radiation creep" effects thus offer opportunities for precision-shaping of metals, semiconductors and ceramics at low temperatures, where thermal activation ($k_B T_{\text{room}}$=0.025eV) plays a minor role compared to energetic radiation (PKA kinetic energy on the order of keV). Also, unlike the temperature which is always a scalar, the radiation stimulus is clearly vectorial as shown in this and previous work (26). Different from traditional swelling and creep models, the "generalized radiation swelling" can have shear components, whereas the "generalized radiation creep" can have hydrostatic strain as well as shear components not parallel to $\sigma_{\text{pre}}$. The tensorial responses open up new routes for reaching the optimal residual stress fields through systematic control of ion/electron radiation energy, beam momentum and focal area, thereby achieving 3DPET. On the other hand, possible mechanical failures caused by shear stresses should be fully considered in applications, as discussed in Supplementary Note 3.

The coupled tuning of momentum and anisotropy stimuli is verifiable and useful. For example, recently, Wei et al. (54) have utilized momentum directionality to generate a stress field and control the grain boundary migration with a focused electron beam in a transmission electron microscope. Furthermore, Rutherford backscattering experiments in channeling directions also detect much deeper backscattering signals than that observed in binary collision simulations in irradiated samples (55, 56). Both ion-beam channeling and focusing may be strongly affected by the longer-range lattice-plane distortions caused by residual stresses, which have not been carefully modelled so far and may be used for verifying the residual stress distribution.

In summary, we proposed a unified framework to describe the relations between residual stresses and radiation PKA stimuli, with a particular focus on radiation-induced residual shear responses. We envision that by rationally designing the radiation flux and external strain fields, one can achieve accurate control of the residual stress fields, which offers new opportunities for elastic strain engineering and nanoscale materials shaping with FIB (2). Further research is needed to verify the possibility and discuss related problems, such as defect migration, stress magnitude verification, and in-service stress interactions. Precise and additive irradiation control by 3D Printing of Eshelby Transformations (3DPET) could play a role in future nanomanufacturing, taking advantage of both "generalized radiation swelling" and "generalized radiation creep" effects.

**Acknowledgment**

QJL and JL acknowledge support by the Laboratory Directed Research & Development Program at Idaho National Laboratory under the Department of Energy (DOE) Idaho Operations Office (an agency of the U.S. Government) Contract DE-AC07-05ID145142, and DOE DE-EE0008830.